\documentclass{epl}

\title{Poling effect on distribution of quenched random fields in a uniaxial
relaxor ferroelectric}

\shorttitle{Poling effect on random fields}

\author{Manuel I. Marqu\'es\inst{1} \and Carmen Arag\'o\inst{1}}
\institute{
  \inst{1} Departamento de F\'isica de Materiales C-IV, Universidad Aut\'onoma de Madrid, 28049 Madrid, Spain\\
}

\begin{document}

\maketitle

\begin{abstract}
The frequency dependence of the dielectric permitivity's maximum
has been studied for poled and unpoled doped relaxor strontium barium niobate
$Sr_{0.61}Ba_{0.39}Nb_{2}O_{6}:Cr^{3+}$ (SBN-61:Cr). In both cases the
maximum found is broad and the frequency dispersion is strong.
The present view of random fields compensation in the unpoled sample is not suitable for explaining
this experimental result.
We propose a new mechanism where the dispersion of quenched random electric fields, affecting
the nanodomains, is minimized after poling.
We test our proposal by numerical simulations on a random field Ising model. Results obtained are in agreement
with the polarization's measurements presented by Granzow et al. [Phys. Rev. Lett {\bf 92}, 065701 (2004)].
\end{abstract}

Relaxor ferroelectrics were discovered almost fifty years ago \cite{Smolenskii} and they
have recently found a multitude of technical applications \cite{Uchino}. In particular, from the
point of view of optical devices, uniaxial strontium barium niobate (SBN) has been proved to be
extremely useful \cite{Miller,Yue}.
Relaxor ferroelectrics present a transition with temperature from polar to non-polar phase
which is characterized by a broad, frequency dependent maximum of the dielectric
permitivity \cite{Samara}.
Another important feature of the relaxor ferroelectrics
is the existence of locally polar nanoregions well above the Curie temperature $T_{c}$ which persist
until the so-called Burns temperature $T_{B}$ \cite{Burns}, which is close to $350^{o}C$ in SBN \cite{LehnenII}.

In SBN, these properties have been studied taking into
account the inherent disorder of the relaxor \cite{Kleemann, KleemannII} by means of a random
field Ising model (RFIM) \cite{Imry}. The disorder produces quenched local random fields
which interact with local dipoles. The local random fields stabilize the dipoles in a small
nanoregion, leading locally to a nonzero value of the spontaneous polarization even above the Curie
temperature. The use of a RFIM to model uniaxial ferroelectrics has been supported by recent findings
in SBN which point toward the existence of internal fields\cite{Granzow,GranzowII,GranzowIII}.

Based on the RFIM, it is possible to make predictions about critical exponents in SBN. Theoretical
calculations for the RFIM predict small values for the spontaneous polarization critical exponent
$\beta=0.06\pm0.07$ \cite{Newman}. Due to this small value, the decrease of polarization when
increasing temperature
close to the Curie point must be very sharp compared with the
behavior of other systems such as Ising models, where $\beta \simeq 0.325$.

Actually, there has been some controversy about the value of $\beta$ for SBN crystals.
NMR measurements yielded $\beta=0.14$ \cite{Blinc} very close to the one predicted by the
RFIM. However, linear birefringence measurements
in SBN resulted in $\beta \sim 0.35$ \cite{Lehnen}, which is actually closer to the value of
the Ising model. This discrepancy has been recently solved by pointing out that NMR samples where fully
poled, while linear birefringence samples where unpoled \cite{GranzowIV}. Granzow et al. have shown
experimentally how the critical exponent $\beta$ varies from $\beta=0.3$ in the case of a
non-poled sample to $\beta=0.126$ when the sample is fully poled.
This is equivalent to say that polarization decreases faster when increasing temperature in poled SBN than in non-poled SBN.

What is the reason for this behavior? A possible answer came from Ref.\cite{GranzowIV}: Basically,
quenched random fields are compensated by those emerging from charged fractal nanodomain
walls in the unpoled sample. Since there is no random electric field, the behavior of the
system resembles the one of an Ising model. However, when SBN is poled, the compensation of
the electric fields is not so effective due to the larger size of ferroelectric domains. Since
the electric fields do not dissapear, the behavior for the poled sample must be the one corresponding to
a RFIM.

Assuming that quenched random fields are responsible for the relaxor behavior in SBN,
a sharper dielectric peak and a decrease in the frequency dispersion should be expected
for the unpoled sample. To check this hypothesis we have performed dielectric permitivity measurements 
in SBN-61:Cr for poled and unpoled samples.

The SBN-61:Cr unpoled samples were small platelets of thickness $0.8mm<d<1.8mm$ and areas between $12mm^{2}$ and $48mm^{2}$.
They were doped with a small amount ($0.01 wt\%$)  of $Cr^{3+}$, but it seems that the impurity type is more
relevant from the optical than the ferroelectric point of view. The two faces perpendicular to
the ferroelectric axis were covered with electro-conductive silver paint.
The samples have been grown by Czocharalski technique at the General Physics Institute of the RAS, Laser Materials and 
Technologies Research Center, Moscow, Russia.
The measurements of the dielectric constant were performed under $1V$ of applied voltage, with a LCR meter HP 4284A whose frequency
ranges from $20 Hz$ to $1MHz$. The temperature control was made through a Unipan thermal controller type 680,
provided with a sensor whose accuracy is $0.1K$ and the sample actual temperature was obtained with a
T thermocouple and a DMM Keithley 196. The poled sample was measured using the following procedure: Once the system reached
a temperature higher than the critical temperature, we pre-polarized
the sample applying a constant $2.8 kV/cm$ DC field while freezing the sample until RT
and then the dielectric constant measurements were performed for the same frequencies that in
the unpoled case.

The experimental measurements presented in Fig.1 show the typical relaxor frequency and temperature
dependence of dielectric constant, with a rounded peak and a non-well defined transition temperature. 
So actually, both samples behave in a very similar way from the relaxor behavior point of view.
The value of the peak of dielectric constant we find for SBN-61:Cr is much smaller than the one found in Ref.\cite{Dec}
for SBN-61:Ce. This small value may be due to the quality of the sample or to the fact that $Cr$ doping increases the SBN
relaxor behavior. In any case, both materials present relaxor properties for poled and unpoled samples.

\begin{figure}
\onefigure {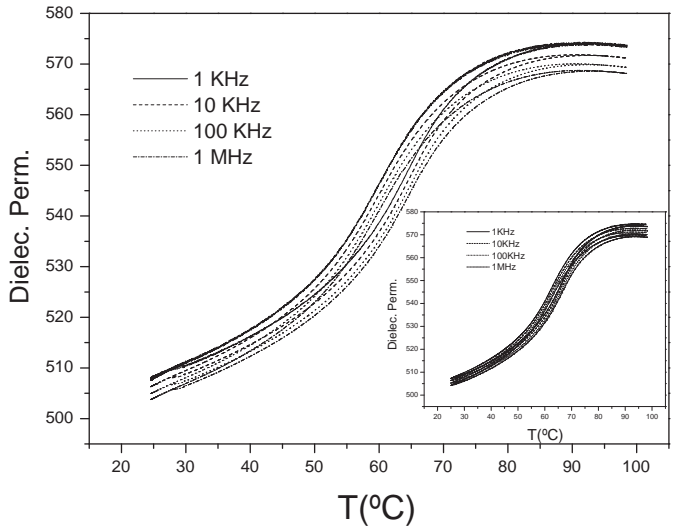}
\caption{Dielectric permitivity vs. temperature for non-poled sample of SBN-61:Cr. Inset: poled sample
}
\label{fig1}
\end{figure}

It is not easy to explain the unpoled results when considering random fields compensation in the
unpoled sample. The broad maximum and the frequency dispersion found for the non-poled sample imply that
quenched random fields are not compensated. On the other hand, the broad maximum also appears for the poled sample,
indicating that random fields are present too. However,
there must be some differences between the fields distributions when comparing both samples,
in order to explain the changes observed in the
polarization behavior \cite{GranzowIV}. How to
explain the dielectric permitivity experimental results and how to make them compatible with the ones observed for the
behavior of spontaneous polarization versus temperature in Ref.\cite{GranzowIV}?

In the following we will propose a model to answer this question.
Local fields are due to positive charges either in deficit at unoccupied $Ba^{2+}$ 
and $Sr^{2+}$ sites in the open tungsten bronze structure \cite{Glass}, or in excess at $Cr^{3+}$ ions
replacing $Nb^{5+}$ \cite{Woike} .

The fields are quenched due
to the low mobility of the charge carriers below the Curie temperature \cite{Buse} and they
are almost uncorrelated due to the short-range character coming from a large dielectric dc
susceptibility \cite{KleemannII}.
An electric field applied to polarize the SBN sample at temperatures below the transition point
should interact with the charge carriers, introducing some degree of correlation between them \cite{DecII}, and
decreasing the value of the dispersion of local fields. Once the electric field is turned off,
the new "ordered" internal electric fields become quenched at room temperature. 
Of course, a change on the dispersion $\sigma$ of the Gaussian distribution of fields in a RFIM
does not change the critical behavior of the system (only the critical temperature) \cite {Newman} so, a
change on $\beta$ is not expected. However, we should take into account that nanodomains found
in relaxors at high temperatures are not formed by just a single dipole. It is necessary
to consider that a quenched randon field affects not just to a single dipole but
to a local group of dipoles forming a nanodomain.
A change on the dispersion of local fields affecting complete nanodomains may change the behavior
with temperature of the total polarization of the system.

We propose the following hypothesis: (i) A poled sample presents a distribution of nanodomain electric
fields with an small dispersion. Due to this small dispersion, all the nanodomains behave similarly
when the critical temperature is approached and they decrease their polarization
almost simultaneously. This commom behavior implies a sharp transition for the total polarization 
versus temperature ($\beta$ decreases). (ii) A non-poled sample presents a distribution of
nanodomain electric fields with a huge dispersion. When the critical temperature is
approached the behavior of the nanodomains is very different from one to the other
and the local polarizations
do not decrease simultaneously. This behavior results on a smooth behavior of the total polarization
versus temperature ($\beta$ increases). However, in both cases, nanodomain electric fields are not compensated 
explaining the broad peak and
frequency dependence found in experiments for both, poled and unpoled samples.

To check this hypothesis we present Monte Carlo simulations of several random field Ising models,
with a Gaussian  distribution of fields centered in zero
and with a width of the distribution given by $\sigma$. We will consider different values of $\sigma$,
corresponding to poled and unpoled samples of
SBN. To study the effect of a random field affecting not just to a single dipole but to a group of
dipoles, we divide the system in several nanoregions $\alpha$, assigning a single electric field to all
dipoles belonging to the same nanoregion. To make calculation easier, we consider a cubic system formed
by $N^{3} \times L^{3}$ dipoles with periodic
boundary conditions formed by small cubic-shaped nanoregions with $N^{3}$ dipoles.
The Hamiltonian reads as follows:

\begin {equation}
H=-J\sum_{<ij>} {s_{i}s_{j}} -J_{h}\sum_{\alpha}\sum_{i\in\alpha} {h_{\alpha}s_{i}}
\label{eq1}
\end{equation}

and the distribution of quenched fields is given by,

\begin {equation}
P(h_{\alpha})=\frac {1} {\sqrt{2\pi}\sigma} \exp \bigl[ {-\frac {h_{\alpha}^{2}} {2\sigma^{2}}} \bigr]
\label{eq2}
\end{equation}

The distributions of fields considered in this work for the poled and the non-poled samples are shown
in Fig.2

\begin{figure}
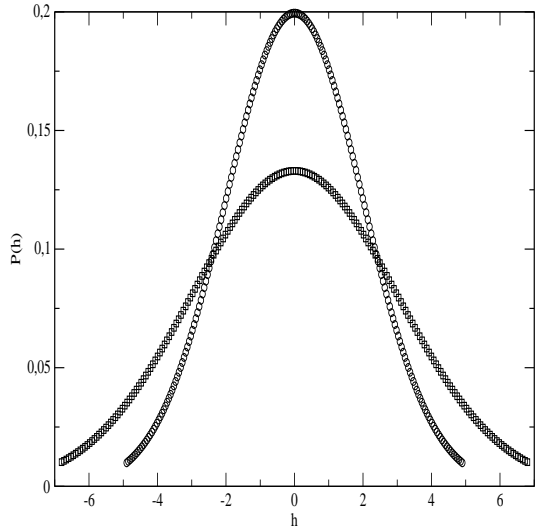

\onefigure[width=7cm,height=7cm,angle=0]{fig2.eps}
\caption{Distribution $P(h_{\alpha})$ of fields for $\sigma=2$, corresponding to the poled sample (circles)
and for $\sigma=3$ corresponding to the un-poled sample (squares).
}
\label{fig2}
\end{figure}

For our simulations we set the values $N=4$, $L=25$, $J=1$ and $J_{h}=0.02$.
In Fig.3a we present results for the total normalized polarization $P=[1/(N^{3} \times L^{3})]\sum_{i} {s_{i}}$
versus temperature for $\sigma = 2$ (representing the poled sample) and $\sigma = 3$ (representing
the unpoled sample). We have fitted the data to a power law
$P(T)=P_{0}(1-T/T_{c})^{\beta}$ for $T<T_{c}$. The values obtained are $T_{c}=4.45$ and $\beta=0.17$
for $\sigma=2$ and $T_{c}=4.44$ and $\beta=0.26$ for $\sigma=3$. These results for $\beta$ compare with the
ones obtained experimentally for SBN in ref.\cite{GranzowIV} for a poled sample (with initial spontaneous
polarization $P(20^{o}C)\sim 10 (\mu C/cm^{2}))$ and a non-poled sample respectively. Fig.3b presents
the power law fits in log scale.

There is a difference between the model we are proposing and the former RFIM studied in detail by Newman and Barkema \cite{Newman}. 
In the former RFIM each single dipole was influenced by a different random field and a change on $\sigma$
affected to the randomness of the system but did not affect to the critical behavior. In contrast, in the present model, the 
same random field affects to a group of $N \times N \times N$ dipoles forming a nanoregion. These nanoregions may act almost
independently from each other (as a collection of non-random Ising models) 
if the differences on the values of the fields are large (i.e. if $\sigma$ is large), changing the critical behavior
from RFIM to Ising model. The value we find for $\beta$ in the case of a poled sample is very similar to the one observed 
experimentally yet not exactly equal to the value ($\beta < 0.06\pm0.07$) reported
for the RFIM in Ref.\cite{Newman}, but it would become very similar in the special case of considering very small nanoregions ($N=1$).

\begin{figure}
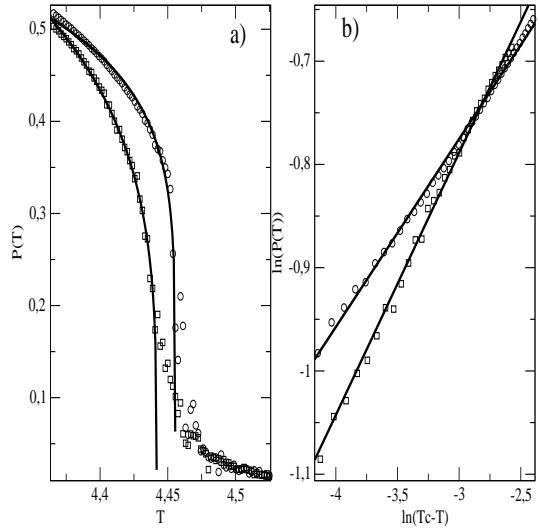

\onefigure[width=7cm,height=7cm,angle=0]{fig3.eps}
\caption{(a) Polarization $P(T)$ vs. temperature $T$ for $\sigma=2$, corresponding to the poled sample (circles)
and for $\sigma=3$ corresponding to the un-poled sample (squares). Continuous lines represent a fitting
to a power law. (b) $ln(P(T))$ vs. $ln(T_{c}-T)$ for $\sigma=2$, corresponding to the poled sample (circles)
and for $\sigma=3$ corresponding to the un-poled sample (squares). Continuous lines represent a linear
fitting.
}
\label{fig3}
\end{figure}

It is also possible to understand this difference on the critical behavior of the polarization by considering
 the
polarization of each single nanodomain versus the temperature. Results are presented in Fig.4 for $50$
different nanodomains. Note how, in both cases, the polarization for each nanodomain remains different
from zero at $T>T_{c}$, as usual for a relaxor. However, the dispersion on the polarization
curves for the poled $\sigma =2 $ case (Fig.4a) is smaller than the one we find for the unpoled $\sigma =3$
case (Fig.4b). The critical temperature for each nanodomain may be determined by considering the
maximum of the derivative of the polarization in each nanoregion versus the temperature. A seasonal
derivative with a period of ten points is shown in Fig.4a and Fig.4b. For a poled
sample, all nanodomains present their critical temperatures in a narrow window around the critical
temperature of the system. Since all nanodomains experiment the transition at almost the same temperature, the
behavior of the spontaneous polarization is very sharp and the value of $\beta$ very small. On the other hand,
when the sample is non-poled, the critical temperatures of the nanodomains are distributed on a wider
temperature range. This makes the total polarization not to decrease abruptly,
obtaining a larger value of $\beta$ than the one corresponding to the poled sample. Relaxor behavior is expected
in both cases, since electric fields are never compensated.

In Ref.\cite{GranzowIV} samples are polarized at room temperature, but it is also posible to polarize at high
temperature and to perform field cooling (FC) \cite{GranzowII}. In this special case, the relaxor exhibits an stable 
polarization at the cooling process with zero field applied. This experimental results are explained by Granzow et al. as 
follows: At high temperatures electric charge carriers are very
mobile and they are able to travel through the crystal and compensate the local random electric fields not in accordance 
with the externally applied electric field. This process imprints an stable non-zero internal field capable of polarizing
the sample at low temperatures. We have measured the critical exponents from the experimental results presented for the SBN 
relaxor at Ref.\cite{GranzowII} and we have found a value of $\beta=0.15$ for the heating process and a value $\beta=0.25$
for the cooling process. Our proposal is supported also by these results obtained for FC samples. From the point
of view of our model, the field compensation coming from the travelling charges at the FC process implies a non-centered 
gaussian distribution of fields (a total value of the internal electric field different from zero). Also, the relaxor 
will show a short range 
ordering as the one already described for the poling process at room tepratures \cite{DecII}, turning intto a decrease on the value of 
$\sigma$ and the value of $\beta$ ($\beta=0.15$). Once the system is heated to very hight temperatures, the internal
electric field still remains \cite{GranzowII}, but the short range ordering dissapears ($\sigma$ turns broad again). Then
,at the cooling process with no external field applied, the system will show the spontaneous polarization comming from a non-centered
distribution of internal fields, with a broad value of $\sigma$, turning into a large value of $\beta$ ($\beta=0.25$). 
Note how, in this case, it is imposible to explain this large $\beta$ value by a total compensation of the internal
fileds (pure Ising model), since non-zero internal fields are needed to explain the existence of spontaneous
polarization when cooling.

\begin{figure}
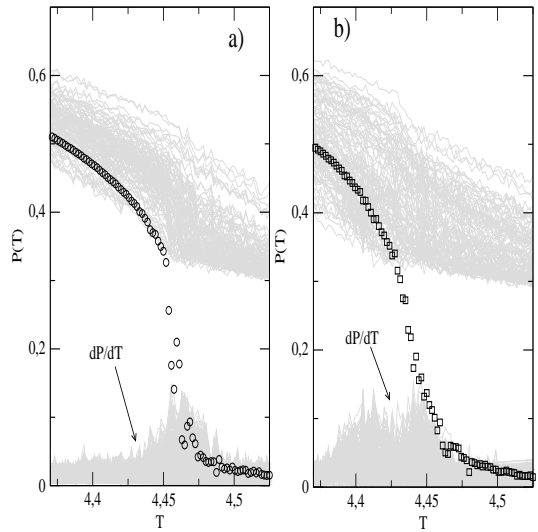

\onefigure[width=7cm,height=7cm,angle=0]{fig4.eps}
\caption{Polarization $P(T)$ vs. $T$ for (a) $\sigma=2$, corresponding to the poled sample (circles)
and (b) $\sigma=3$, corresponding to the non-poled sample (squares).
Grey lines represent polarization of each nanodomain ($P_{\alpha}$) and
$dP_{\alpha}/dT$ vs. temperature for $50$ different nanodomains.
}
\label{fig4}
\end{figure}

In summary, the dielectric dispersion in relaxor SBN is similar for the non-poled and for
the poled sample. For this reason, an explanation for the change on the critical
behavior of polarization based on fields compensation in the unpoled sample is not appropriate.
A mechanism capable of explaining both behaviors has been proposed.
In this new mechanism a poled sample presents
an smaller dispersion of quenched electric fields than a non-poled sample. Monte Carlo
simulations show how this model leads to a different behavior of the local polarization in the
nanodomains of the relaxor system. This difference on behavior produces a change on the
criticality of the total polarization similar to the one found experimentally.

\acknowledgments

We thank M.Ramirez and J.Garc\'ia Sol\'e for supplying the samples. Helpful discussions with
J.A.Gonzalo and W.Kleemann are greatfully acknowledged. This work was supported by the DGICyT
through grant BFM2000-0032.

\end{document}